# Astro-animation - A case study of art and science education

**Laurence Arcadias**
Maryland Institute College of Art, Baltimore.
larcadias@mica.edu

**Robin H.D. Corbet**
University of Maryland, Baltimore County; X-ray Astrophysics Laboratory, NASA Goddard Space Flight Center; CRESST; Maryland Institute College of Art.
corbet@umbc.edu

**Declan McKenna**
Maryland Institute College of Art, Baltimore; NASA Goddard Space Flight Center
dmckenna@mica.edu

**Isabella Potenziani**
Maryland Institute College of Art, Baltimore; University of Maryland, Baltimore County; NASA Goddard Space Flight Center, CRESST
bpotenziani@mica.edu

**Abstract**
Art and science are different ways of exploring the world, but together they have the potential to be thought-provoking, facilitate a science-society dialogue, raise public awareness of science, and develop an understanding of art. For several years, we have been teaching an astro-animation class at the Maryland Institute College of Art as a collaboration between students and NASA scientists. Working in small groups, the students create short animations based on the research of the scientists who are going to follow the projects as mentors. By creating these animations, students bring the power of their imagination to see the research of the scientists through a different lens. Astro-animation is an undergraduate-level course jointly taught by an astrophysicist and an animator. In this paper we present the motivation behind the class, describe the details of how it is carried out, and discuss the interactions between artists and scientists. We describe how such a program offers an effective way for art students, not only to learn about science but to have a glimpse of "science in action". The students have the opportunity to become involved in the process of science as artists, as observers first and potentially through their own art research. Every year, one or more internships at NASA Goddard Space Flight Center have been available for our students in the summer. Two students describe their experiences undertaking these internships and how science affects their creation of animations for this program and in general. We also explain the genesis of our astro-animation program, how it is taught in our animation department, and we present the highlights of an investigation of the effectiveness of this program we carried out with the support of an NEA research grant. In conclusion we discuss how the program may grow in new directions, such as contributing to informal STE(A)M learning.

**Introduction**

Animation has the power to bring things to life in a unique light (Amidi 2017). It enables us to tell stories at an emotional and metaphorical level that can touch everybody in a universal way. Astrophysics also addresses fundamental questions of human existence. The night sky is part of the shared heritage of all people on Earth (Venkatesan & Burgasser 2017), creating a sense of belonging deeply rooted in our collective ancestors' memory (Tingay 2015). This feeling of wonder and notion of place in the universe connects us physically and emotionally through a personal experience to the rationale of science. For example, the philosopher Merleau-Ponty commented that - "all my knowledge of the world, even my scientific knowledge, is gained from my own particular point of view, or from some experience of the world without which the symbols of science would be meaningless" (Merleau-Ponty 1962)**.** The Square Kilometre Array, which will be the largest radio telescope in the world, spanning from Australia to South Africa, is a beautiful example of how science can learn from the ancestral knowledge of Indigenous people to build a stronger understanding of the sky shared around the globe. In the case of the SKA, art was used as a channel leading to *Shared Sky,* a travelling collaborative art exhibition with Indigenous artists from Australia and South Africa celebrating ancient and modern stories about the cosmos (Mann 2016).

When integrated with science, animation becomes a powerful tool with a friendly touch that may help break science's intimidating aspects (e.g. Dalacosta et al. 2009, Barak et al. 2011). Since the origin of cinema, animation and astrophysics established an early relationship.

The first science-fiction movie: *A Trip to the Moon* (Méliès 1902) was more of a trick film[1] but did include the first almost-animated sequences. *The Einstein Theory of Relativity* (Fleischer 1923) attempted to make an educational film for a non-scientific audience. Many notable animation films would follow, either as fiction, documentary, or educational movies, making science an inspirational topic to be explored. This is masterfully demonstrated in *Of Stars and Men* (John and Faith Hubley 1964), a feature-length animated film based on the book by scientist Harlow Shapley. The New York Times noted in 1964: "At no time is his illustration so cute or coy or inept that it fails to convey precisely his intellectual point".

    Animation theorists have noted the role and importance of animation in documentaries through film history (Strøm 2001). Wells (2016) argued that the animated documentary in its own right has only recently been accepted as proven and valid. This was also established through Roe's book: *Animated Documentary* (2013) which noted that Folman's animation film *Waltz with Bashir* (2008) took a step away from children and family audiences with a serious topic about lost memories following trauma. The series *Cosmos A Spacetime Odyssey* (Braga et al. 2014) uses animation to convey historical aspects of astronomy. In the very first *Cosmos* series *A Personal Voyage* (Sagan 1980), Blinn, the famous computer graphics pioneer, created an animation on evolution, using metamorphosis as a continuous line-drawing representation of species in transformation.

    Here we describe a specific example of a collaboration between animation and astronomy in the form of a course taught at an art school in collaboration with NASA scientists. This paper presents our astro-animation program, summarizes the main conclusions of our research into its benefits and outcomes, and discusses its potential for further growth and development.

**Astro-Animation Program at MICA**
The Maryland Institute College of Art (MICA) is a private art and design college located in Baltimore with a fast-growing undergraduate animation department. Along with fundamental animation courses, the curriculum offers a variety of elective project classes. One of the most popular is the astro-animation class which is conducted in collaboration with scientists located at the NASA Goddard Space Flight Center (GSFC), a major research centre located near Washington DC.

    The astro-animation project started from a shared passion for astronomy and animation. MICA is only 40 minutes away from NASA GSFC which facilitates informal connections. Hence, introducing animation students to astronomy research during Arcadias' advanced animation class was an appealing project to try out. After a discussion with Corbet about his research at NASA GSFC, he introduced the project to his colleagues in the Fermi satellite team

---

[1] Trick film refers to live effects including stopping the camera to produce an illusion while animation film simulates movements synthetically through a series of pictures.

and several scientists volunteered to participate as mentors. The result was successful and evolved into our astro-animation class (Arcadias & Corbet 2015).

The students get a chance to work closely with astronomers to create animations based on the scientists' research. Even if the power dynamic seems unbalanced, with undergraduate students only starting an introductory course in astrophysics, while the NASA scientists are experts in their fields, the relationship still works as an exchange of talents. The animation students, even if often initially intimidated, are excited to offer their vision, while most scientists are eager to have their research re-imagined. The students learn to work as a team and to communicate with somebody who has no familiarity with the world of animation, while scientists have to clearly explain their research to be understood. The scientists are also curious to learn more about the animation process and have asked the students to explain their work techniques, which is included in the final presentations of the films.

The program has several goals. For the animation component, students learn to use scientific concepts as a source of inspiration and to explore different ways of thinking about animation outside the box. For the science part, the pedagogic outcomes require that students understand basic concepts of astrophysics and the possibilities and limitations of scientific research more broadly. By working directly with scientists, the aim is that they will learn that science is at least as much a way of thinking as it is a body of knowledge (e.g. Sagan 1995). Another goal is to underline the importance of the artist's point-of-view on science, and having students offer their own artistic response to scientific topics. The work can be whimsical or poetic but nevertheless constrained by scientific rigor. Our approach is different from the visualizations that NASA generally produces, mostly graphic illustrations of scientific results, depicting astronomical phenomena as "accurately" as possible. While the students' films are free-form animated responses, perhaps more aligned with Roe's (2013) description of the animated documentary: "Animated documentaries offer us an enhanced perspective on reality by presenting the world in a breadth and depth that live-action alone cannot". Students also learn how to translate scientific concepts into a meaningful or intriguing piece of animation able to reach out to a broader public. Animation is a powerful way to present scientific research to a broad audience. Halas and Batchelor (1981) stated that "If it is the live-action film's job to present physical reality, animated film is concerned with metaphysical reality - not how things look but what they mean" (Wells 1998).

**Class Organization**
The astro-animation class is jointly taught by an astrophysicist (Corbet) and an animator (Arcadias). It takes place for 15 weeks with three credits awarded for science and three credits for animation. The class starts with astronomy and then transitions to an animation studio. Hands-on experiments and demonstrations during pre-COVID-19 time (Figure 1) helped students connect with their own art making. The animation instruction includes presentations on artists inspired by astronomy, and how the cosmos has been represented through history and across diverse cultures.

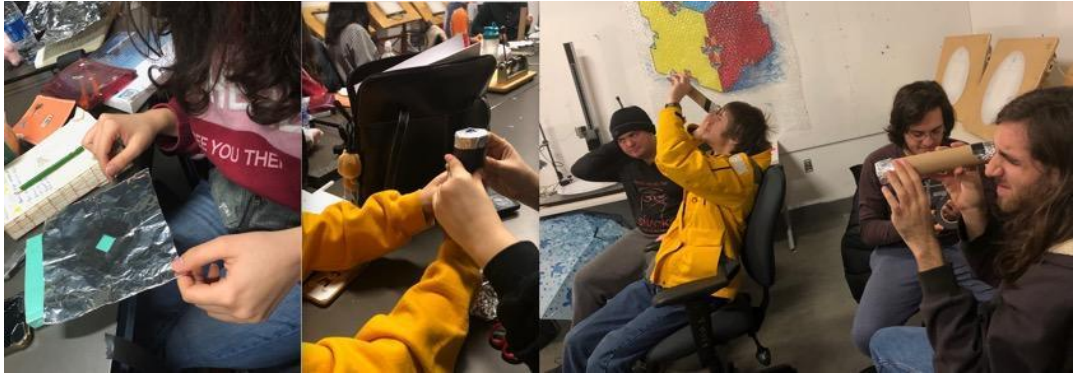

*Figure 1: Science experiments and demonstrations: Studying light with spectroscopes constructed by the students.*

During the first two weeks, students hone their animation skills on broad scientific concepts such as gravity and the scale of the Universe. It pushes them to connect spontaneously with their personal experiences to represent the "unrepresentable" aspects of the universe (Figures 2 and 3).

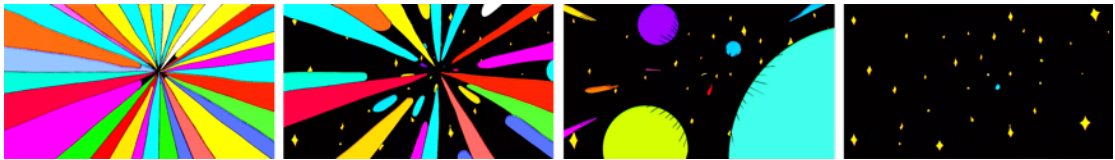

*Figure 2: Sneeze - Tenzin Lhamo*

*I interpreted the Big Bang to start as an unprompted moment that has forever changed the universe. A concept much easier to grasp is a sneeze. A person sneezes which causes germs, to shoot out of one's mouth in any direction…*

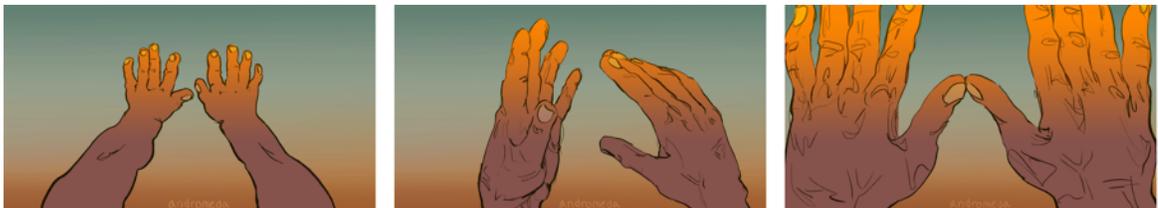

*Figure 3: Andromeda - Terri Ogwulumba*

*When reading about the scale of the universe, and realizing the relationship between distance and time, I found it fascinating that our sight of the Andromeda Galaxy is of something millions of years in the past. (...) it is interesting to consider the idea of light as a messenger and the idea that an object like my hand is stuck a little bit more in the past if I hold it far away from my eyes than if I hold it close.*

In week three, the scientists come to the class to present their current research topics such as black holes, supernovae, dark matter, and gravitational waves (Figure 4). Students are randomly assigned a topic and a mentor scientist (Figure 5). They work in groups of two or three to produce approximately six to eight projects in total. Communication between the student teams and their mentors happens through email and social network sites such as Tumblr (Figure 6).

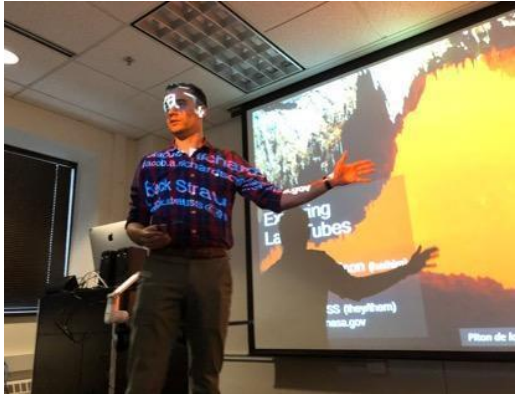
*Figure 4: Scientist presenting his research*

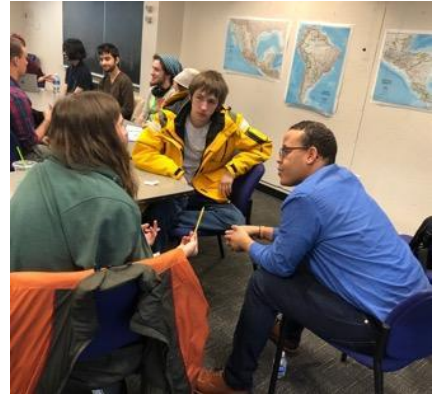
*Figure 5: Students talking with a scientist*

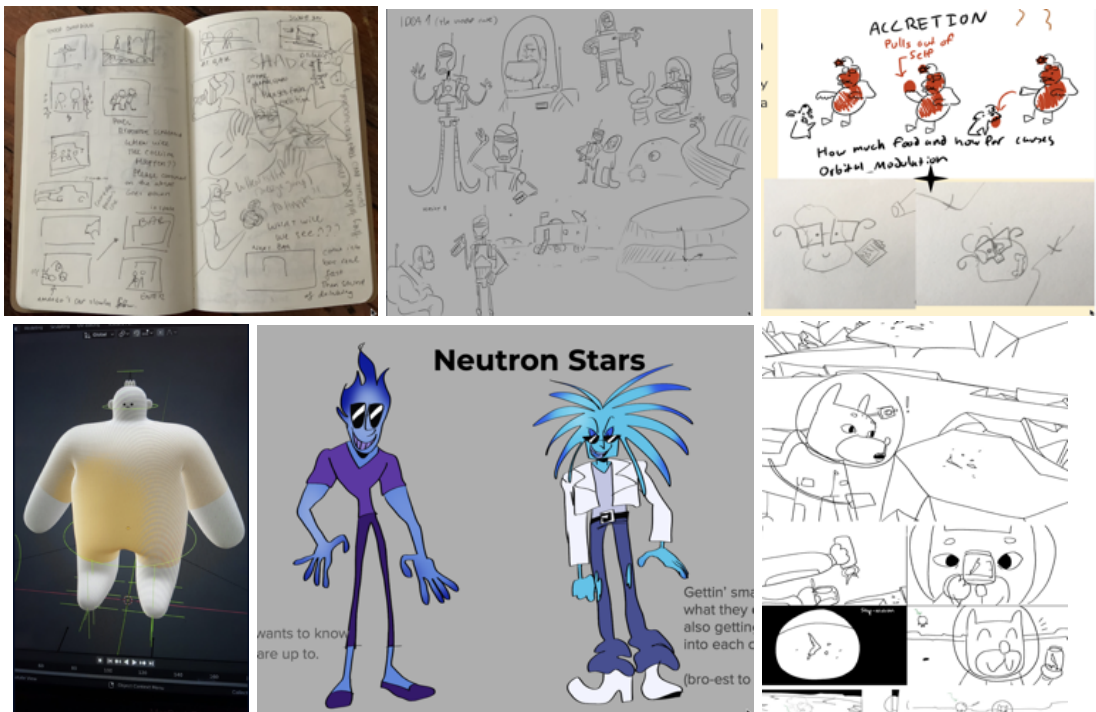
*Figure 6: Examples of graphic research from various student projects.*

In week five, students visit NASA GSFC (Figure 7a). After a tour of the campus, they present their concept designs, storyboards and animatics (Figure 7b) to the scientists during a lunch-time session (Figures 8a, b).

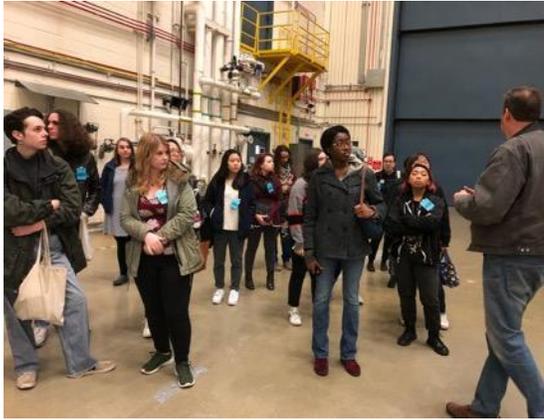 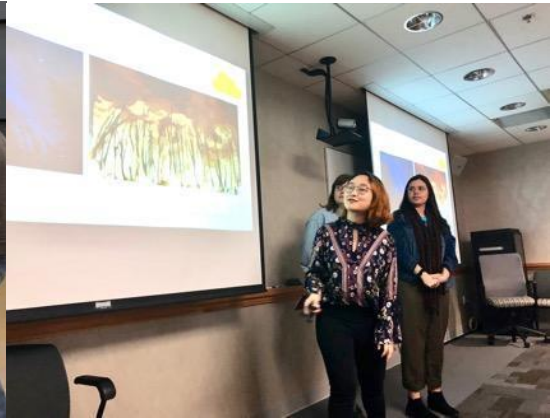

*Figure 7a: Students touring NASA GSFC*    *Figure 7b: Students presenting storyboards*

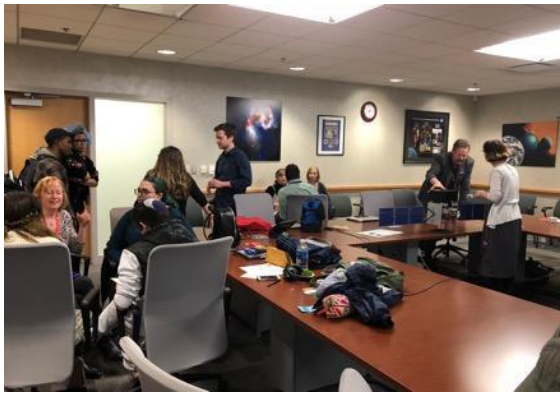 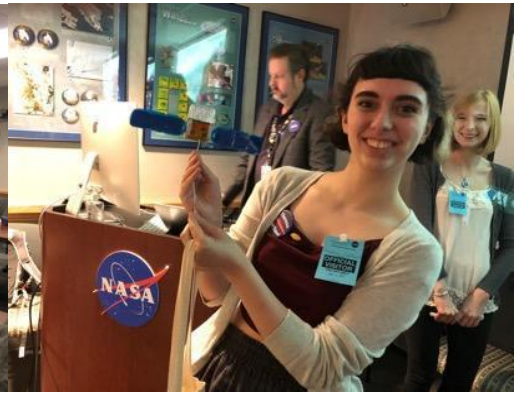

*Figure 8a: Post-presentation discussions*    *Figure 8b: Student showing her Fermi satellite model*

During weeks six to 13, students work on the production of their animations for a final screening at the NASA GSFC Visitor Center. The screenings have occurred on "Take Our Daughters and Sons to Work Day" (Figure 9b). The latest live screening in 2019 had a large audience that included many employees with their children, which we have to keep in mind for our presentation. Even if many of the animations the students produce are child friendly, they are not specifically aimed at children and the target audience is the students' decision.

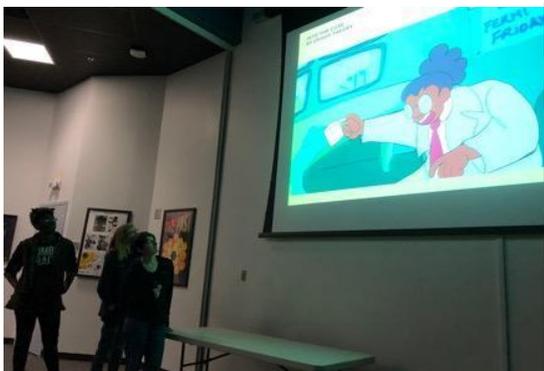 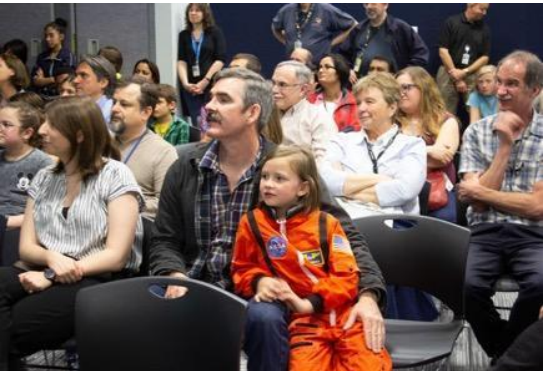

*Figure 9a: Students presenting their animation*    *Figure 9b: Take Our Daughters to Work Day at NASA GSFC*

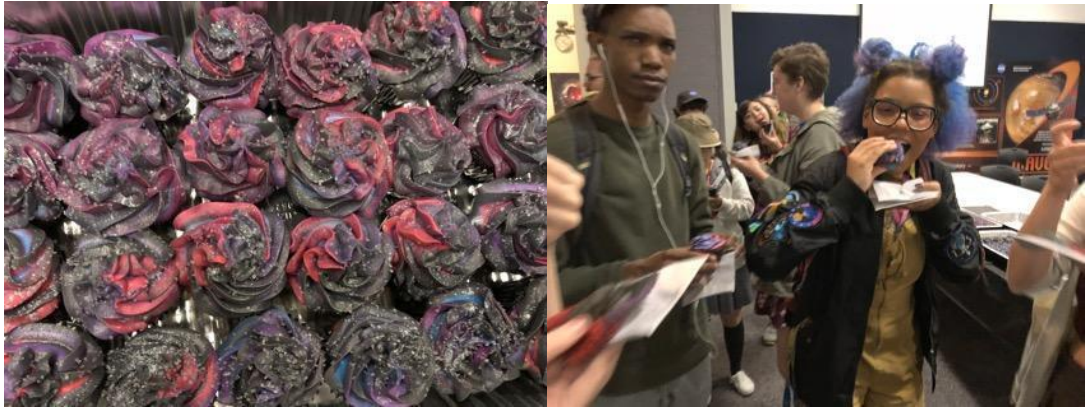
*Figure 10: Galaxy cupcakes by the Fermi cake committee*

Every year, the Fermi Cake Committee has generously rewarded the students with their own space-inspired cake creations and is now part of the tradition (Figure 10). The Fermi Gamma-ray Space Telescope team at GSFC have been valuable supporters from the very beginning. After this screening, the animations are finalized for our website - AstroAnimation.org. As of early 2021, there are forty-seven animations covering a range of topics. Each film is typically about two minutes long and is freely available for download under a licence that allows non-commercial use. The animations have been featured at a variety of additional venues including science fiction conventions, art festivals, a STEAM[2] festival, public outreach presentations, and have been used in college and university science classes. They have been presented in a variety of formats, including "stand-alone" events hosted by scientists and animators, or only scientists (Figures 11, 12).

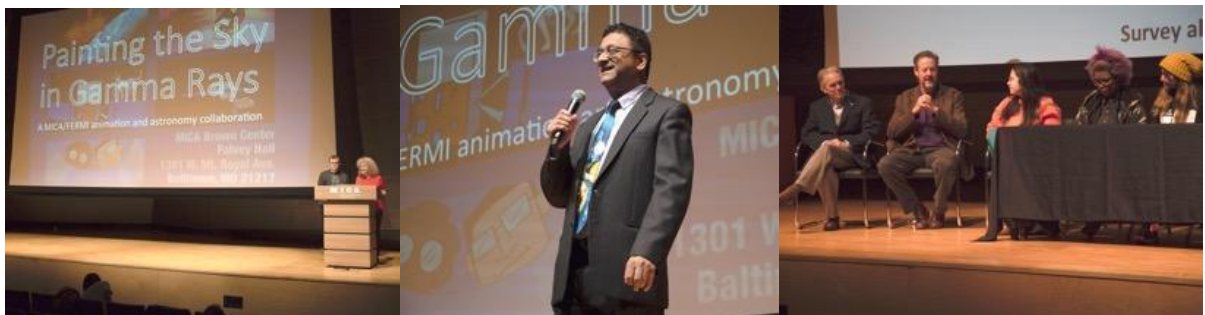
*Figure 11: "Painting the Sky with Gamma Rays" screening at MICA curated by scientists and animation students during the 8th Fermi symposium- Oct 2018*

---

[2] STEAM: Science, Technology, Engineering, Art and Mathematics. A for Art was added later to STEM.

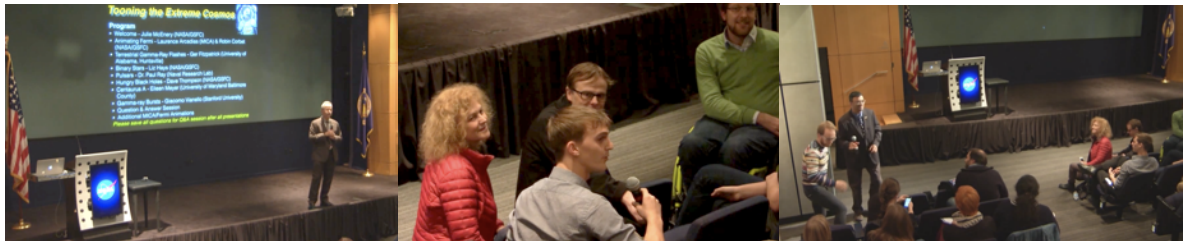

*Figure 12: Paul Hertz Director of the Astrophysics Division presenting the "Tooning the extreme cosmos" screening at NASA Headquarters in Washington DC and a MICA student answering questions - Fermi Symposium 2015*

**Artistic Implications**

**Metaphoric or poetic interpretations in the animations**

Animation has the power to tell stories in a variety of styles and approaches that are unique to this medium (Wells 1998). In *Understanding Animation*, Wells talks about narrative strategies specific to animation. When the students explore a scientific concept and turn it into an animation it is expressed in a totally different light from the raw science. It takes a new form that can be playful, surprising, humorous or metaphoric and become more accessible to some viewers (Figure 13).

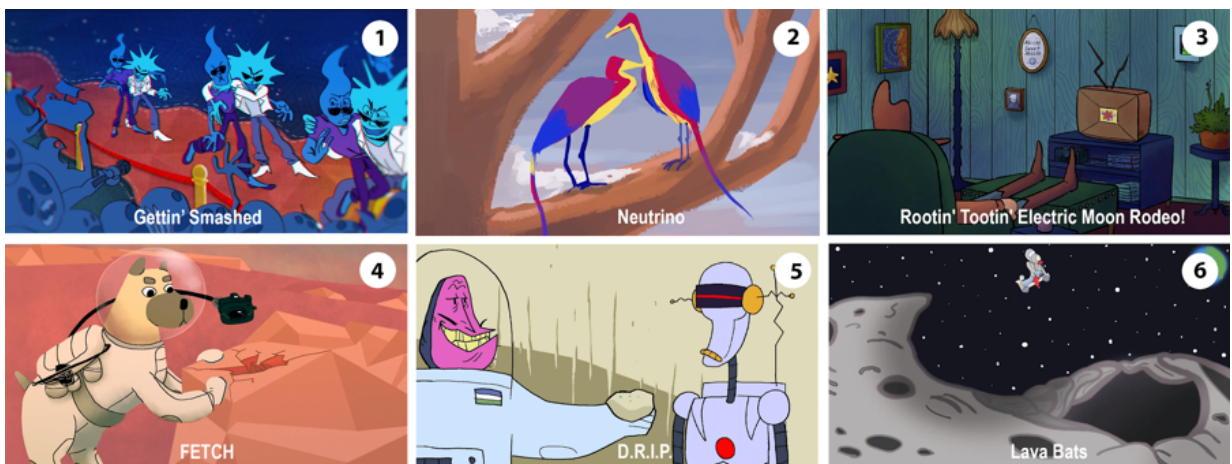

*Figure 13: Various metaphorical representations*

*#1- Paparazzi tail celebrated neutron stars during bar crawling. #2- A documentary inspired approach depicting a neutrino, a gamma ray, and protons as birds taking a journey. #3- An astronaut traveling in lunar craters as a cowboy trying to find electricity to power his Moon barn. #4- The Martian sequel (on Titan) with a dog replacing Matt Damon. #5- A search for water on the Moon with a human astronaut versus a robot. #6- Bat scientists investigating lunar lava tubes, inhabited by a mysterious cult.*

**Character Representation**

Over the seven years that the class has been taught, the animations from the students have evolved. One early issue was to break clichés of scientists in character representation. By bringing scientists into the classroom, students realized that a lot of scientists looked like them, although typically older, from various ethnic origins, and even having similar personalities. This change of perception has been reflected in recent animations through more playful representations of scientists (Figure 14).

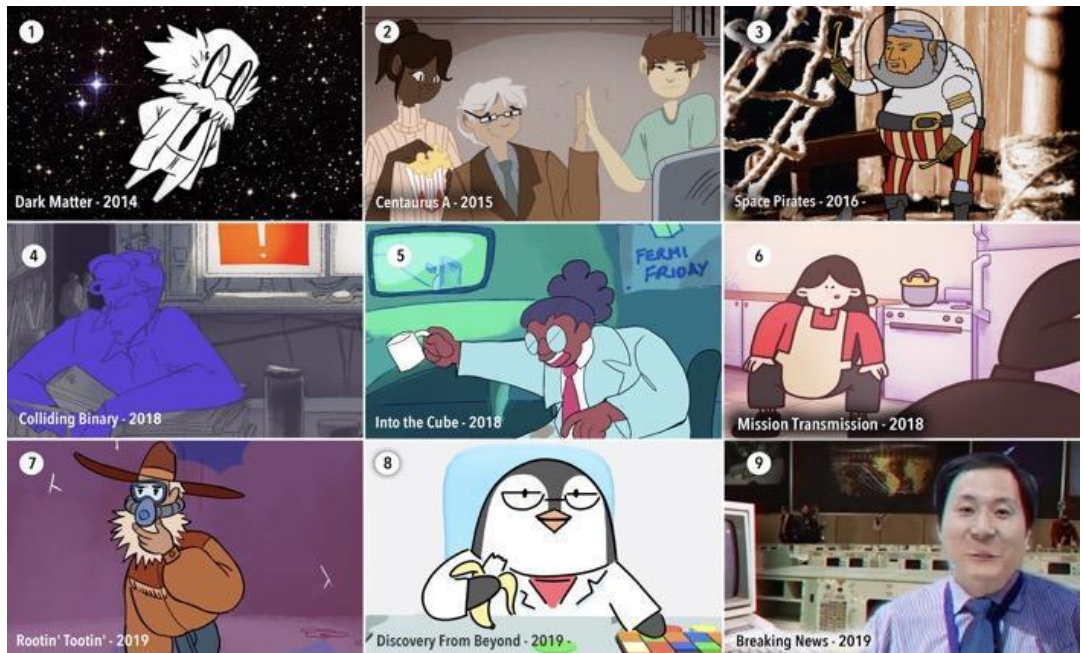

*Figure 14: In these images from student films from 2014 to 2019, we see the evolution of the representation of scientists.*

*# 1, scientist as an Einstein, white male figure. # 2, more inclusive ethnic, age and sex representations but still in traditional roles with the older man in control. # 3, 7, 8, artistic licence with scientists as a pirate, a cowboy, and an anthropomorphized animal. # 6 represents a mom scientist in both her role of parent and scientist. # 4 and 5 portray scientists with borderline behaviours (sleeping at work, low key). # 9 is an Asian-American scientist from the '80s/'90s. # 5 is an African American female scientist. # 1,2,5,9, scientists shown wearing ties.*

In addition to scientist figures, the animations often depict other characters including children and their parents in family situations. We noted that to date no father characters have been portrayed while mother figures are often represented, as well as non-traditional family models such as a single mother or lesbian parents (Figure 15). We speculate that students no longer feel the need to have a father or male figure for validation. The female figure, either as a mother or a scientist, has the same authority and conveys identical trust. The racial diversity of the students in the classroom is well represented through the multiple ethnicities of the characters, including the scientists, showing a level of comfort from our students to identify with these characters (Figure 14).

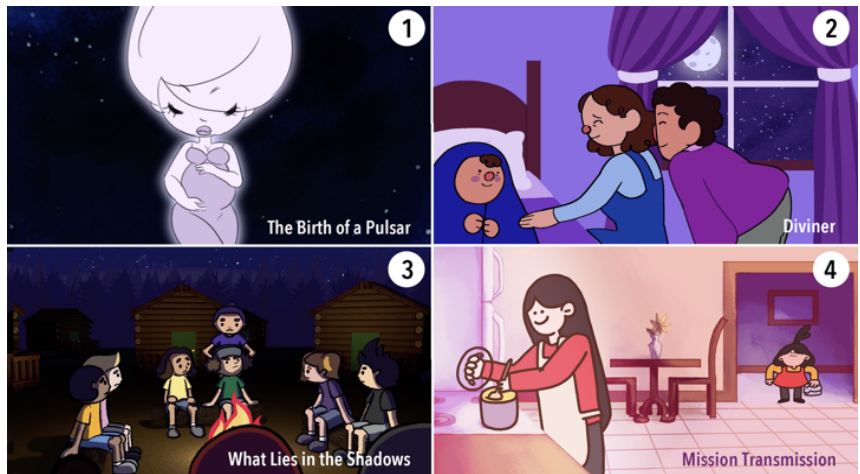

*Figure 15: Various parental representations*

# 1, A mother star pregnant with her baby neutron star. # 2, One family included a small child and her two same-sex parents. # 3, One mother character in a group of children around a campfire. # 4, One single mother with her daughter.

**Techniques**

One goal is to push students to be as experimental as possible using any technique they want, from 2D to 3D to stop-motion and to explore their personal vision (Figure 16).

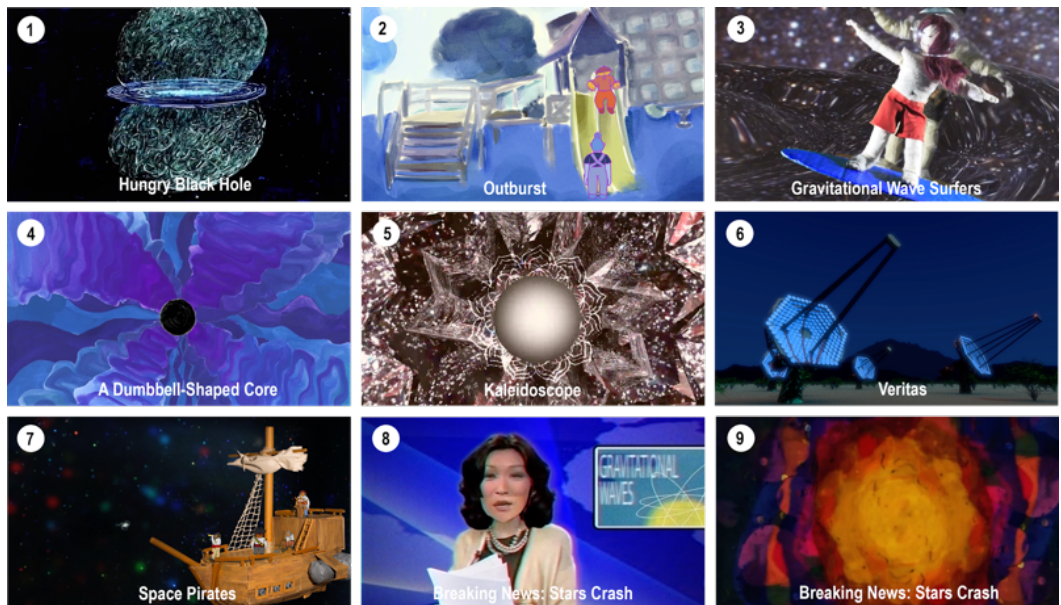

*Figure 16: Various animation techniques used by students*

# 1- Paint on glass. # 2- Watercolour. # 3- 3D+ stop-motion. # 4- Cut out paper.
# 5- 3D. # 6- 3D. # 7- 2D and stop-motion. # 8- Digital cut-out. # 9- Wax strata-cut.

**Evaluation of the astro-animation project**

We received a National Endowment for the Arts (NEA) grant to evaluate the benefits and outcomes of the astro-animation project. We conducted surveys and interviews among a variety of groups, including scientists, animation students, and non-specialist audiences. The overall conclusion was that presenting astronomy via animation is highly attractive to a range of audiences. We also found that simply screening the animations on their own has a visual impact. However, for more specific science learning it is valuable to have some additional scientific presentation.

**Scientific accuracy and artistic licence**

One question we investigated is how far we can go with the visual representation of the scientific concepts and how to give as much creative freedom as possible to the students while maintaining scientific accuracy. Most scientists have been very open to students' interpretations. For example, a student represented the inside of a black hole as a donut world into which an astronaut was swallowed (Figure 17). The scientist's reaction was that since we don't know the answer, any interpretation is fine, underlining that science is as much about what we don't know as what we do know. This opens a door for artistic exploration, potentially engaging with scientific dilemmas. But some scientists may not have the same level of comfort with the students' creative licence, as reflected in a comment from the survey that the science may get lost, or miscommunicated. When the animations try to be educational, there is a greater chance the students may misrepresent scientific facts, while inspirational pieces are more open to interpretation. A complex science problem may be difficult to understand in a few minutes of animation, but instead, the animation may help "engage with the science and suggest different ways of seeing" (Ede 2005). It can also inspire the audience to look for more information.

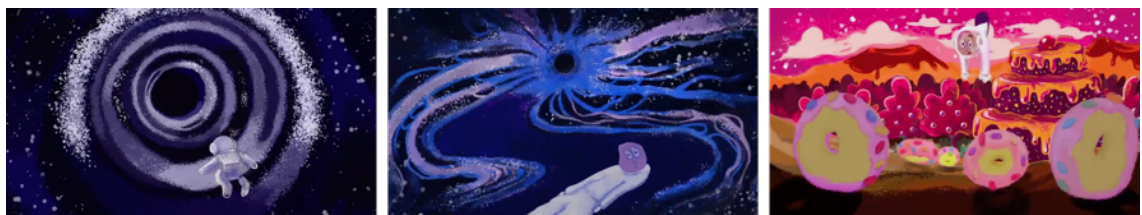

*Figure 17: Donut Hole- Puglisi, Whang, Wang '15*

**Art and Science as Mutual Inspiration**

We also explored whether this collaboration has led to any impact from art on the scientific research being undertaken. Some authors, such as Root-Bernstein (1999), Miller (2001, 2014), and Chomaz (2018) have considered potential influences by art on scientific research. For example, Root-Bernstein (2000) states that many scientific techniques originate from art. Anamorphosis, a Renaissance perspective technique, helped Thompson and Huxley describe evolutionary and embryological processes as anamorphic distortions. As a side note, for an animator it is exciting that astronomers discovered Pluto using a basic animation technique of flipping between two photographs of the same sky area obtained at different times[3]. Lehrer (2008) noted that "science needs art to frame the mystery, but art needs science so that not everything is a mystery. Neither truth alone is our solution, for our reality exists in plural". Similarly, Shlain (1991) reported that "artists have mysteriously incorporated into their works, features of a physical description of the world that science later discovers". He also addresses the parallels between the works of Picasso and Einstein on time and space without any direct interaction between the two, even though they both revolutionized art and science around the same time.

      Within the scope of our program, the answer to how art can challenge science is still unclear at this level. In our surveys and interviews there is little evidence of the art students affecting the actual research. However, the experience forced the scientists to clarify their thoughts for better communication to students and consequently other non-specialists. In the MICA students' interviews, even if this is certainly not a class requirement, a few felt and stated that their animations had to be didactic: "There is a limit, you have to do something educational", "visual explanations are really the key to explaining things to the public". These comments made us aware that the students felt they needed permission to push animations in the direction they wished. Perhaps an upper-level astro-animation class could have more potential to stimulate art exploration, leading to some form of scientific discovery or deeper art/science intersection.

      Another critical component of the class is to expose art students to the scientific method. It is important to know that science is a process, a way of questioning the world constantly, or as philosopher Karl Popper put it, scientific knowledge "consists in the search for truth, but it is not the search for certainty... All human knowledge is fallible and therefore uncertain" (Popper 1996). Working with scientists may give animation students a better understanding of the scientific process and help them establish a method of their own. Scientific research and art research share some similarities, emotion and intuition can even play an important role in science. "All great achievements of science must start from intuitive knowledge… At times I feel certain I am right while not knowing the reason" Einstein confided (Calaprice 2000). Being exposed to these differences and similarities between science and art practice may inspire students to further investigate the world through different approaches.

**Student Internship Experiences at NASA**
Each year, an astro-animation student has spent the summer at NASA GSFC for a paid internship to work with scientists and outreach professionals. In addition, several students have had the opportunity to undertake internships for credit. In summer 2020, two students worked as interns but, because of the COVID-19 pandemic, had to undertake this from home. Here are the reports of their experiences.

---

[3] PlutoBlink GIF: https://www.sightsize.com/wp-content/uploads/2015/08/PlutoBlink-Animated.gif

**Isabella Potenziani:**
As an animator who works in stop-motion and 2D, I was intrigued to see where my style of animation and astrophysics could lead to. With the mentorship of Corbet and Jay Friedlander, I was able to learn more about the astrophysics division, as well as new animation skills. I worked with many other NASA scientists, one of them, Roopesh Ojha, always took the time to explain and go over information related to projects. Everybody was very patient with their explanations providing me with resources that helped inform my projects.

    Despite some challenges due to COVID-19, the overall experience was enriching and engaging. I produced several projects I never imagined I would get the chance to create. Two of the smaller projects I worked on during my internship were the making of a 2D GIF and a 3D model in Maya. I had never done any intricate modelling, especially nothing based on a specific object (Figure 18 #1). For the 3D model of Surveyor 1-7, I used only photos and a few technical drawings. It was a different experience making something that accurately represented all the parts of the craft. It was difficult at times considering many of the photos were taken from similar angles.

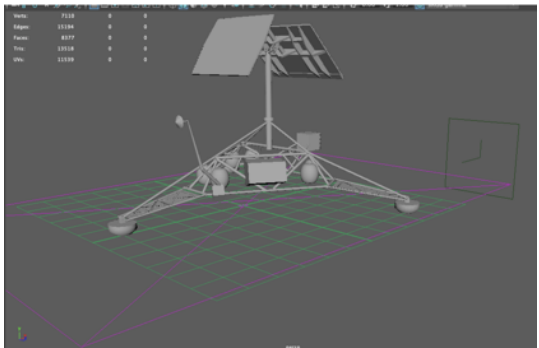 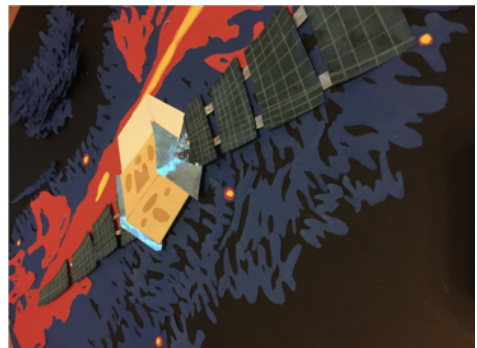

*Figure 18:  #1-3D model of Surveyor 1-7              # 2- Fermi's 12th anniversary*

    My largest project was the Fermi satellite's 12th anniversary animation, using paper cut-out puppets animated and overlaid onto a hand-cut paper background in post-production (Figure 18 #2). I really enjoyed working with several scientists in the division to create this piece. I also was pleased to be making it in stop motion. It was challenging at times with the limitations I had, but it did help me learn new techniques and ways of doing stop motion.

**Declan McKenna:**
My time as an Undergraduate Research Associate in Astrobiology focused on animation as a method of and response to inquiry in astrobiology. Previous experiences with astronomical topics prepared me for the program, such as my time assisting Arcadias, and the previous creation of a film in the astro-animation course. At GSFC, I explored how to make animations that are purposefully engaged with astrobiology.

Two questions were asked: what is the role of animation to the science of astrobiology, and what is the impact of scientific research on the practice of animation?

I began with a series of interviews with animators, lab scientists, and media writers. Guided by Andrea Jones, one of GSFC public engagement leads, I found that animation is most useful in scientific contexts when it aids in understanding for an audience and displays the scientific process and alternative perspectives on the research. As well, animations must be built from trust and knowledge of the subject, and be appropriate for the way they are being presented. The next phase of the research was emphasizing these qualities through direct practice.

To facilitate an intentional approach to my animations[4], I developed the *Scientific Method of Animation*. Similar to the scientific method, this process incorporates five steps:

1. Ask a question that animation will answer.
2. Create an objective, such as the intended final output.
3. Experiment with the creative and research processes.
4. Draw conclusions from feedback.
5. Adjust, and arrive at a finished piece.

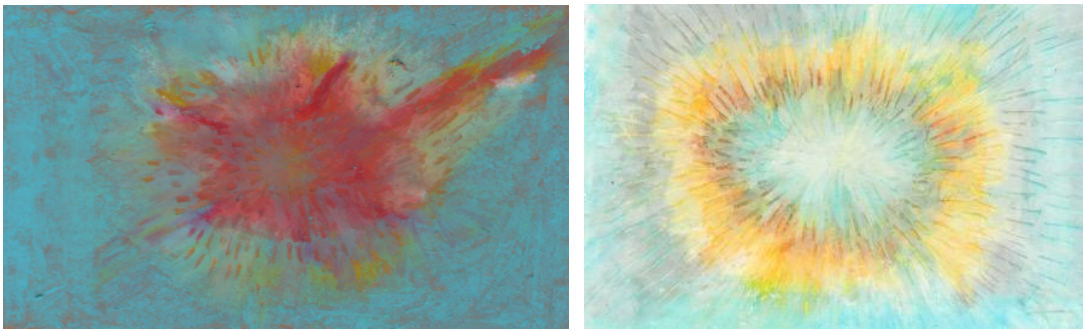

*Figure 19: #1 Stills from two abstract gifs, created using oil pastels, depicting the glow of meteors as they enter Earth's atmosphere (left) and #2 the natural processes that occur inside of asteroids, such as water alteration (right).*

This methodology was used for a NASA Web Feature on the Asuka 12236 Chondrite. Under Lonnie Shekhtman, Senior Science Writer at Goddard, the question "What is the relation of Asuka to the themes of life?" was posed, and the objective of two editorial gifs established. After researching Chondrite studies in the analysis of early-Earth biological development, several thumbnails and style tests were pitched. I made adjustments based on the feedback and created two hand-drawn, abstract gifs interpreting Asuka, that were subsequently published (Figure 19).

---

[4] Usually, animation is not defined as a method, it's up to the animator to choose or potentially create one. We note that Nelson (2013) specifies a process which has some similarities to the one described here.

Operating remotely complicated my work - in-person meetings present opportunities to draw, discuss, and restructure material alongside the scientists, and this process is difficult to translate online. However, the discussions, and conferences I attended engaged me in the NASA world and led to several incredible collaborations, such as an augmented reality depiction of Ocean Worlds in our solar system (Figure 20). These projects showed that my animation directly benefits from a methodical and intentional ideation process, like those governing the scientific process, to increase accessibility. Compared to surface-level engagements, a work environment similar to that of a scientific institution, one that focuses on immersion with research, collaborators, and experimentation, leads to more introspective and informed approaches to artistic creation.

Astronomical research and animation both investigate humans' unique relationship to the universe and to being alive. Integrations between art and science help ground and ignite passion between researchers, artists, and the audience, connecting all parties together in the exploration of the meaning of life.

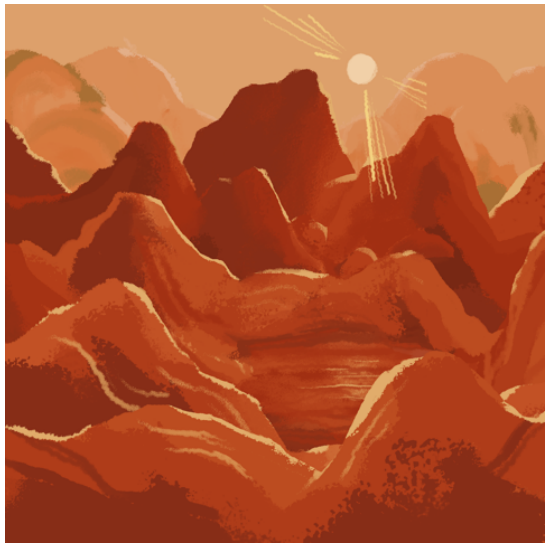

*Figure 20: Illustration of the surface of Titan, used for an Augmented Reality Instagram filter.*

**Conclusion and Future Directions**

Overall, we found the collaboration between animation students and NASA scientists to be highly effective and well-received by most participants. The program was found to be a successful way to engage art students with science. At the same time, a large number of scientists have found it worthwhile to participate and have been excited to see their scientific research converted into an animation. The contributions from the intern students are very important for them to learn how they can find their place as artists working around scientists. Potenziani managed to do a stop-motion piece for the Fermi gamma-ray space telescope's 12th anniversary animation which was very successful and unexpected since graphic work at NASA is usually digital. The made-by-hand cut-out technique added a human-touch warm feeling to the birthday celebration of the satellite. McKenna worked in a different division and describes his own animation process as a parallel to the scientific method that could be understood by his team. That was an efficient way to highlight the importance of his role as animator when investigating astrobiology for outreach.

In addition to continuing the program, we are exploring ways to develop it and take it outside of the classroom. One promising new avenue would be to use astro-animation as a basis for an informal learning program. In a "Guerrilla-STEM"[5] setup, we intend to design an itinerant astro-animation exhibition with hands-on artistic and scientific activities, to be displayed in unexpected locations such as libraries, train stations or music festivals. The goals are to engage participants who are not a priori attracted to, or aware of science. Integrating animation with astronomy, both of which have general popularity, may generate unique appeal, curiosity and initiate easier learning. The long-term benefits of this project would include increased interest in STE(A)M, expanded participation in science by underrepresented groups, and enhanced interdisciplinary collaboration in science and art.

**Acknowledgements**
We thank many people at NASA and elsewhere who have contributed to this project, all the students who have participated, and the paper referees for valuable comments. This work was supported in part by the NEA (17-3800-7013) and NASA (80GSFC17M0002).

**References**

Arcadias, L. & Corbet, R. (2015), 'Animating Fermi - A Collaboration Between Art Students and Astronomers', *Leonardo*, 48, no. 5.

Amidi, A. (2017), '3 Key Ways That Europe's Feature Animation Scene Differs From The U.S.', *Cartoon Brew*. https://www.cartoonbrew.com/feature-film/3-ways-europes-feature-animation-scene-differs-u-s-150081.html

Barak, M., Ashkar, T. and Dori, Y.J. (2011), 'Learning science via animated movies: Its effect on students' thinking and motivation'. *Computers & Education*, 56(3), pp.839-846.

Botticello, Casey (2019), https://caseybotticello.medium.com/what-is-guerilla-art-91c7dfbcbea5. Accessed 31 January 2021.

Braga, Brannon, Bill Pope, Ann Druyan, Livia Hanich, Steve Holtzman, Steven Soter, Alan Silvestri, et al. (2014), *Cosmos: a spacetime odyssey*.

Calaprice, A. (2000), *The Quotable Einstein,* Princeton University Press; 2nd edition

---

[5] Guerrilla STEM derives from Guerrilla Art: "A method of art making where the artist leaves anonymous art pieces in public places" (Botticello 2019). Similarly, the Guerrilla Science group (Zivkovic 2011) employs "'science by stealth' … to stretch traditional boundaries of how people engage with science".

Chomaz, J.M. (2018), 'It Is Time to Think the Anthropocene! A Manifesto,' *Leonardo*, Volume 51**,** Issue 2, 217-219.

Dalacosta, K., Kamariotaki-Paparrigopoulou, M., Palyvos, J.A. and Spyrellis, N. (2009), 'Multimedia application with animated cartoons for teaching science in elementary education'. *Computers & Education*, 52(4), pp.741-748

Ede, S. (2005), *Art & Science*, London: I.B.Taurus

Fleischer, David (1923), *The Einstein Theory of Relativity,* USA: Fleischer Studios.

Folman, Ari (2008), *Waltz with Bashir*, Israel: Bridgit Folman Film Gang.

Halas and Batchelor from <Hoffer, T. 1981, Animation: a Reference guide, Westport Connecticut: Greenwood Press> in Wells, P., 1998

Understanding Animation London: Routledge, p11

Hubley, John and Faith (1961), *Of Stars and Men,* USA: Brandon Films.

Lehrer, J. (2008), *Proust was a Neuroscientist*, New York: Mariner Books.

Mann, Adam (2016), 'Science and Culture: Looking at a shared sky, through the lens of art', *Proceedings of the National Academy of Sciences*, 113(50), pp.14165-14167.

Merleau-Ponty, Maurice (1962), *Phenomenology of perception*. London: Routledge.

Masetti, M. (2014), 'The Nexus of Art and Science, Part 2', *Blueshift.* https://asd.gsfc.nasa.gov/blueshift/index.php/2014/08/25/can-you-animate-the-sound-of-a-gamma-ray-burst/ Accessed 31 January 2020.

Méliès, G. (1902), *A Trip to the Moon*. France: Blackhawk Films.

Miller, A. I. (2001), *Einstein, Picasso: Space, Time, and the Beauty That Causes Havoc*. New York. Basic Books.

Miller, A. I. (2014), *Colliding Worlds: How Cutting-Edge Science Is Redefining Contemporary Art*. New York. W.W. Norton & Company, Inc.

Nelson, R. (2013) *Practice as Research in the Arts*, Houndsmills: Palgrave Macmillan.

Popper, Karl Raimund (1996), *In Search of a Better World: Lectures and Essays From Thirty Years*. New York: Routledge.

Roe, A.H. (2013), *Animated documentary*. New York: Palgrave Macmillan.

Root-Bernstein, Robert and Michèle (1999), *Sparks of Genius,* New York: Mariner Books.

Root-Bernstein, Robert (2000), 'Art advances science', Nature, Vol 407, 14 September.

Sagan, Carl, Ann Druyan, Steven Soter, Adrian Malone, Tom Weidlinger, Geoffrey Haines-Stiles, David Kennard, et al. (1980), *Cosmos: a personal voyage*. Studio City, CA: Cosmos Studios.

Sagan, Carl (1995), *The Demon-Haunted World: Science as a Candle in the Dark*, Ballantine Books.

Shlain, Leonard (1991), *Parallel Visions in Space, Time, and Light,* New York: William Morrow & Co.

Snyder, Timothy (2021), 'The American Abyss', *The New York Times Magazine*, 9 January.

Strøm, Gunnar (2001), 'The animated documentary, a performing tradition', [*Norsk medietidsskrift 02 /*(Volum 8)](#)

Tingay, S. (2015), 'Indigenous culture and astrophysics: a path to reconciliation', *The Conversation*, July 6.

Venkatesan, A. & Burgasser, A. (2017), 'Perspectives on the Indigenous Worldviews in Informal Science Education Conference', *The Physics Teacher*, 55, 456.

Wells, P. (1998), *Understanding Animation*, London: Routledge.

Wells, P. (2016), 'Writing Animated Documentary: A Theory of Practice'. *International Journal of Film and Media Arts*, *1*(1).